\documentclass[12pt,a4paper]{article}

\makeatletter
\@addtoreset{equation}{section}
\makeatother

\newcommand{\bsb}{$\mbox{MKK} \mbox{--} \overline{\mbox{MKK}}$}
\begin{document}
\begin{flushright}
\footnotesize
\footnotesize
CERN-TH/2000-352\\
Bicocca-FT-00-23\\
\normalsize
\end{flushright}

\begin{center}

\vspace{.8cm}
{\LARGE {\bf Branes from Unstable Systems of Branes
\footnote{To be published in the proceedings of the RTN-meeting 
{\it The Quantum Structure of Spacetime and the Geometric Nature
of Fundamental Interactions}, Berlin, October 2000.}}}

\vspace{1cm}


{\bf Laurent Houart}

\vspace{.1cm}

{\it Dipartimento di Fisica\\
Universit{\`a} Milano Bicocca\\
Piazza delle Scienze 3, Milano, Italy}\\
{\tt Laurent.Houart@mi.infn.it}

\vspace{.3cm}

{and}

\vspace{.3cm}

{\bf Yolanda Lozano}

\vspace{.1cm}

{\it Theory Division, CERN\\
1211 Gen\`eve 23, Switzerland}\\
{\tt yolanda.lozano@cern.ch}

\vspace{.4cm}

\vspace{1cm}


{\bf Abstract}

\end{center}

\begin{quotation}

\small

We discuss various aspects of the description of branes as
topological solitons in unstable brane systems of higher 
dimensions. We first describe a classification of all the possible
realisations of branes of M and type II theories as topological 
solitons of a brane-antibrane system. We then present a description
of type IIB NS-NS $p$-branes in terms of topological solitons in 
systems of spacetime-filling NS9, anti-NS9 pairs and discuss the 
implications of these constructions in the description of BPS and
non-BPS states in the strongly coupled Heterotic SO(32) theory.
We finally present briefly the construction of a conjectured 
spacetime-filling non-BPS M10-brane, starting point for a brane descent
construction of the branes of M-theory.

\end{quotation}

\vspace{1cm}

\newpage

\pagestyle{plain}

\section{Introduction and Summary}

Recently, significant progress has been made
in our understanding of M-theory beyond the BPS configurations.
These advances have already permitted to test the web of dualities 
relating the different phases of M-theory  (the different superstring 
theories) on some of the  non-BPS states of the spectrum.  
A beautiful outlook on the interplays between BPS branes
and non-BPS branes has also been given in some cases (see
\cite{rev} for reviews and references therein) along with
an elegant mathematical formulation \cite{kteo,kteo2,hora2}. 
For example, a BPS  Dp-brane in type II may be viewed as coming
from a non-BPS system given by a D(p+2), anti-D(p+2) pair.
The instability of this non-BPS configuration manifests itself in a 
complex tachyonic mode of the open string stretched between the pair. 
When the pair coincides, the
tachyon can roll down to a true vacuum  and condenses in a topologically 
non-trivial way leading to a stable  
vortex-like configuration, and the resulting object is
a BPS Dp-brane \cite{senb}, \cite{sena}.

Most of the progress has been made in the context of D-branes.
It would certainly be interesting to understand if similar constructions
exist for the other $p$-branes like for instance the M-branes or
the NS-NS charged ones. We will describe here some aspects of this
problematic.
We will first discuss a systematic classification of all
the possible realisations of branes of M and type II theories as
topological solitons of brane-antibrane systems. This classification 
of all the possibilities, consistent with the structure of the theory,
is achieved by studying the Wess-Zumino terms in the worldvolume
effective actions of the branes of M-theory and their reductions.
We will illustrate the method on the {\bsb} system.
We will then turn to a description of the brane descent construction
in type IIB starting with a spacetime-filling brane-antibrane system
NS-NS charged and discuss the implications of this construction in
the description of BPS and non-BPS states in the strongly coupled 
heterotic SO(32) theory. Finally we will construct the worldvolume action 
of a conjectured spacetime-filling non-BPS M10-brane, starting point for
a  brane descent construction in M-theory. 
This talk is based on \cite{HL1}, \cite{HL2} and
\cite{HL3} where further details can be found.

\section{Realisation of branes as topological solitons}

The (bosonic) low energy effective worldvolume theory of a Dp, anti-Dp pair 
is described by two U(1) field strengths, one $F=dA$  
for the brane and one $F^{\prime}=dA^{\prime}$ for the
antibrane, along with a complex tachyon field charged under the relative
U(1). The D(p-2) brane is obtained as a topological soliton of the
system after the tachyon condensates, rolling down to its true 
vacuum in a topological non-trivial way, giving rise to a usual
vortex solution \cite{sena}, \cite{senb}. 
The fact that this codimension two object carries
a D(p-2) brane charge can be deduced from the following Wess-Zumino
term in the action of the pair:
\begin{equation}
\label{senex}
\int_{\Sigma_{p+1}} C^{(p-1)} \wedge (F - F^{\prime})
\end{equation}
which indicates, after integration on the codimensions, that
the vortex is charged under the RR-field $C^{(p-1)}$ and corresponds
thus to a D(p-2) brane.

Having understood the construction for Dp-branes and using the
Web of dualities relating the different phases of M-theory, it
is possible to qualitatively discuss all the possible realisations of 
branes of M and type II theories as topological solitons in
brane-antibrane systems. The classification of all the possibilities
is achieved by studying the
Wess-Zumino terms in the worldvolume effective actions of the branes of
M-theory and their reductions. We will here illustrate the method
by analysing the Kaluza-Klein, anti Kaluza-Klein monopole pair 
 ({\bsb}) and refer the reader to \cite{HL1} for 
the other cases and the complete classification.

The worldvolume effective action of the M-theory Kaluza-Klein monopole
was constructed in \cite{BJO,BEL}. The existence of the Taub-NUT 
direction in the space transverse to the monopole
is implemented at the level of the effective action by 
introducing a Killing isometry which is gauged in the worldvolume. 
Then the target space fields must couple in the worldvolume
with covariant derivatives of the embedding scalars, or through 
contraction with the Killing vector. The Kaluza-Klein monopole is charged
with respect to an 8-form, which is the electric-magnetic dual of the
Killing vector considered as a 1-form. 
This field is itself contracted with the Killing vector, giving a 7-form
minimally coupled to the 7 dimensional worldvolume of the monopole.

The worldvolume effective action of the monopole contains the following
term \cite{BEL}:
\begin{equation}
\label{M2deM5}
\int_{R^{6+1}} i_{\hat k}{\hat {\tilde C}}\wedge d{\hat b}^{(1)}\, ,
\end{equation}

\noindent where ${\hat {\tilde C}}$ denotes the 6-form of eleven 
dimensional supergravity, ${\hat k}$ is the Killing vector, with
$(i_{{\hat k}}{\hat {\tilde C}})_{{\hat \mu}_1\dots {\hat \mu}_5}
\equiv {\hat k}^{{\hat \mu}_6}{\hat {\tilde C}}_{{\hat \mu}_1\dots
{\hat \mu}_6}$, and ${\hat b}^{(1)}$ is a 1-form worldvolume field
which describes the coupling to an M2-brane 
wrapped on the Taub-NUT direction. The same coupling appears in the
effective action of the {\bsb} pair, where now 
${\hat b}^{(1)}={\hat b}^{(1)}_1-{\hat b}^{(1)}_2$, with 
${\hat b}^{(1)}_{1,2}$ the corresponding vector field in the worldvolume
of each monopole. This term allows for an object of codimension 2,
associated to the localised magnetic flux which accompanies
the  topologically non-trivial  condensation of the
tachyonic mode of the M2-brane. The integration of the localised flux  
$d{\hat b}^{(1)}$ on the codimensions gives

\begin{equation}
\int_{R^{4+1}} i_{\hat k}{\hat {\tilde C}}
\end{equation}

\noindent i.e. the solution is an M5-brane soliton with one worldvolume
direction wrapped around the Taub-NUT direction. Therefore we can describe
a (wrapped) M5-brane soliton as a bound state of a pair of
Kaluza-Klein anti Kaluza-Klein monopoles.

We can now analyse the different possible configurations in the type 
IIA theory
to which this process gives rise.
Dimensionally reducing along the Taub-NUT direction of the monopoles
we can describe a D4-brane through the condensation of the tachyonic
mode of an open
string between a D6, anti-D6 pair \cite{sena}. 
This is described by the
coupling:

\begin{equation}
\int_{R^{6+1}}C^{(5)}\wedge db^{(1)}
\end{equation}

\noindent which is straightforwardly obtained by reducing the coupling
(\ref{M2deM5}) describing the creation of the solitonic 
(wrapped) M5-brane. 

The reduction along a worldvolume direction of the monopoles gives as
one of the possible configurations a solitonic NS5-brane, 
obtained after the
condensation of the tachyonic excitation of an open string 
stretched between a Type IIA pair of Kaluza-Klein anti-Kaluza-Klein monopoles.
The worldvolume reduction of (\ref{M2deM5}) gives indeed:

\begin{equation}
\int_{R^{5+1}}i_k C^{(5)}\wedge db^{(1)}+\int_{R^{5+1}}i_k B^{(6)}\wedge
db^{(0)}
\end{equation}

\noindent where $b^{(0)}$ arises as the component of ${\hat b}^{(1)}$
along the worldvolume direction that is being reduced. These terms
describe two processes: One in which a (wrapped) D4-brane is created
after condensation of the tachyonic mode of a
(wrapped) D2-brane stretched between two Type
IIA monopoles, described by the first
term, and one in which a (wrapped) NS5-brane is created after the
condensation of a (wrapped) open string.
This is, to our knowledge, the first example in which
a NS5-brane has been described through a brane anti-brane pair
annihilation. 

We now turn to the ``dual'' process in the {\bsb} system  , i.e.  
obtaining an M2-brane soliton after the condensation of an M5-brane. 
A wrapped M5-brane is coupled in the worldvolume of
the MKK through a 4-form field, which is the worldvolume dual of the
vector field describing  the coupling of a wrapped M2-brane.

The Wess-Zumino term of the dual description
of the {\bsb} system was found in \cite{HL1}, and it is given by:

\begin{equation}
\label{dualb4}
\int_{R^{6+1}}(i_{\hat k}{\hat C})\wedge d{\hat b}^{(4)}
\end{equation}

\noindent where $d{\hat b}^{(4)}$ is the worldvolume dual of the center of mass
vector field ${\hat b}^{(1)}_1+{\hat b}^{(1)}_2$.
This term allows for a codimension 5 soliton corresponding to a wrapped
M2-brane. Indeed, after integrating the localised
$d{\hat b}^{(4)}$ flux on the codimensions one obtains:

\begin{equation}
\int_{R^{1+1}}(i_{\hat k}{\hat C})\, ,
\end{equation}

\noindent which is the charge of a wrapped M2-brane.

The reduction along the Taub-NUT direction 
of the coupling (\ref{dualb4}) gives:

\begin{equation}
\int_{R^{6+1}} B^{(2)}\wedge db^{(4)}\, .
\end{equation}

\noindent Therefore, we find after the integration of the 5-form
field strength 
a solitonic object of codimension 5 charged under $B^{(2)}$,  
namely a fundamental string, obtained after the condensation
of a streched D4 between a D6, anti-D6 pair. 
Processes giving rise to fundamental strings where first study in \cite{Yi}. 

Two other possible brane anti-brane annihilation processes can be
deduced in Type IIA 
from the M-theory Kaluza-Klein anti-Kaluza-Klein annihilation
that we have just discussed.
If we reduce this process along a worldvolume direction of the
Kaluza-Klein monopoles we find two terms.
The first term describes a fundamental string
created after the condensation of a NS5-brane stretched between
a pair of Type IIA Kaluza-Klein monopole anti-monopole. 
Both the fundamental string and
the NS5-brane are wrapped on the Taub-NUT direction of the 
monopole. On the other hand, the second term represents a wrapped
D2-brane arising after the condensation of a wrapped D4-brane
stretched between the two monopoles.

\section{Spacetime-filling NS-NS branes and brane descent relations}

In this section, we first discuss the possibility of describing
the NS-NS charged branes of type IIB in terms of a system of 
spacetime-filling  NS9, anti-NS9 pairs \cite{HL2}.

The NS9 is predicted by the spacetime SUSY algebra of the type IIB theory,
and its worldvolume effective action
was constructed in \cite{BEHHLvdS}. 
The explicit form of the Wess-Zumino part describing 
$N$ NS9-branes reads:

\begin{eqnarray}
\label{NS9ac}
&&S_{{\rm NS9}}^{{\rm WZ}}=\int_{R^{9+1}}{\rm Tr}\left[ B^{(10)}+
{\tilde C}^{(8)}\wedge {\tilde F}+B^{(6)}\wedge {\tilde F}
\wedge {\tilde F}+
C^{(4)}\wedge {\tilde F}\wedge {\tilde F}\wedge {\tilde F}
+\right. \nonumber\\
&&\left. +B^{(2)}\wedge {\tilde F}\wedge {\tilde F}\wedge {\tilde F}
\wedge {\tilde F}+\frac{C^{(0)}}{(C^{(0)})^2+e^{-2\phi}}
{\tilde F}\wedge {\tilde F}\wedge {\tilde F}
\wedge {\tilde F}\wedge {\tilde F}\right]\, .
\end{eqnarray}

\noindent Here ${\tilde F}\in U(N)$ describes 
D1-branes ending on the NS9-branes,
$B^{(10)}$ is the NS-NS
10-form potential with respect to which the NS9-brane is charged and
${\tilde C}^{(8)}$ is the S-dual of the 8-form RR-potential 
(see \cite{EL}). 

The effective action describing NS9, anti-NS9 pairs
contains additional worldvolume fields. One has, in addition to ${\tilde
F}$, 
a second field strength ${\tilde F}^{\prime}$, associated with the antibranes,
and a complex charged tachyon field $T$. 
However, to discuss
the possible realisations of p-branes as topological defects of the NS9,
anti-NS9 system 
one can focalise on the WZ terms of the brane action (\ref{NS9ac}),
bearing in mind
that the topologically non-trivial character of the soliton can be 
carried by one of the two field strengths, say ${\tilde F}$ 
(see for instance \cite{sena}).

The analysis of the NS9-brane WZ action shows that the NS-NS branes
of the type IIB theory can be understood as bound states of
NS9, anti-NS9 pairs of branes. The different possible classical 
solutions corresponding to topological
defects of codimension $2k$ can be obtained starting with $N=2^{k-1}$ pairs
of NS9, anti-NS9 using the 
Atiyah Bott Shapiro realisation 
\cite{abs} of  the generator of the homotopy group $\pi_{2k-1}((U(2^{k-1}))$,
as explained in \cite{kteo} for the case of the D9, anti-D9 system.
The general pattern that is derived from this analysis is: D7=(NS9, anti-NS9),
NS5=2 (NS9, anti-NS9), D3=4 (NS9, anti-NS9), F1=8 (NS9, anti-NS9)
and D(-1)=16 (NS9, anti-NS9), of type IIB p-branes as bound states 
of NS9, anti-NS9 pairs. 
Therefore, we see that all the branes predicted by the analysis of
the type IIB spacetime supersymmetry algebra, apart from the pp-wave
and the Kaluza-Klein monopole, can be realised as
bound states of any of the two types of spacetime-filling branes of
the theory. Of course for this to hold the NS9-branes
must be considered on an equal footing with the RR 9-branes.   
The pp-wave and the Kaluza-Klein monopole solutions are only defined
in spacetimes with one special, isometric, direction.
In the case of the pp-wave this is the direction of propagation
of the wave, whereas in the monopole case this is the 
Taub-NUT fiber of the transverse space. Therefore these branes 
cannot be understood as bound states of spacetime-filling branes 
that do not see any of these special directions.

We know turn to the implications of these constructions in the
description of BPS and non-BPS states in the strongly coupled
Heterotic SO(32) theory.

It was argued in \cite{Hull2} \cite{BEHHLvdS} that the Heterotic
string with gauge group SO(32) could be obtained as a nonperturbative
orientifold construction of the type IIB theory. This construction is
determined by S-duality: If the type I theory can be defined as an
orientifold of the type IIB theory by its worldsheet parity reversal
symmetry, its S-dual, i.e. the Heterotic SO(32), should be defined at
strong coupling as an orientifold of the type IIB theory by the
S-dual operation of worldsheet parity reversal. Perturbatively, this
operation coincides with the $(-1)^{F_L}$  symmetry of the
type IIB theory, where $F_L$ is the left-moving fermion number.
Modding out the type IIB theory by $(-1)^{F_L}$ gives
rise to the type IIA string, and this is determined 
by the twisted
sector that has to be added in order to restore modular invariance.
However one can consider adding anomaly cancelling 9-branes 
as in the type I
theory, and this is the way in which the Heterotic string 
can be produced. S-duality
determines that the background 9-branes must be NS9-branes, and also
that, nonperturbatively, $(-1)^{F_L}$ can be defined as the operation
reversing the orientation of a D-string. This worldsheet operation has
associated an orientifold fixed plane with -32 NS-NS charge, and this
charge is cancelled through the addition of 32 NS9-branes, with one
unit of NS-NS charge. Together they
reproduce the gauge sector of the Heterotic SO(32) supergravity
\cite{BEHHLvdS}.

The massless spectrum of the Heterotic
F-string can be described in terms of open D-strings with both ends 
on the F1, DD D-strings,
and with one end on the F1 and the other on an NS9-brane, 
DN D-strings, \cite{BEHHLvdS}.
The DD D-strings contribute with 8 scalars and 8 right-handed 
Majorana-Weyl spinors, whereas the DN D-strings contribute with 32
left-moving fermions. This description
of the Heterotic string arises at strong coupling. However,
given that the massless states are BPS, it can be as well extrapolated
to the weak coupling regime. In this limit the SO(32) charges tethered
to the F1 are pulled onto its worldsheet, since 
$\tau_{{\rm D1}}/\tau_{{\rm F1}}\sim 1/g_s\rightarrow\infty$ and the
D-strings collapse to a point, giving the usual SO(32) Heterotic
worldsheet currents.

So far the discussion has been focussed on the massless BPS states.
The Heterotic SO(32) theory contains  
as well perturbative massive
states in the spinorial representation of SO(32), which are non-BPS
but stable, given that they are the lightest states transforming as
spinors of SO(32). Sen \cite{sena}, \cite{senb} 
showed that the correct way of 
describing these states at strong coupling is in terms of a weakly
coupled type I (D1, anti-D1) system. A (D1, anti-D1) pair is unstable
due to the presence of a tachyonic mode in the open strings stretched
between the two branes, and moreover, since the two branes are
spinors under SO(32) the bound state cannot transform as a spinor.
However, compactifying the D-strings and switching on a $Z_2$ Wilson
line the tachyonic mode can condense into a stable configuration
different from the vacuum, and transforming as a spinor of SO(32)
\cite{sena}. 

We can now take the following point of
view to describe the spinorial non-BPS states of the Heterotic theory
in the strong coupling regime.
A bound state D0=(D1, anti-D1) in the weakly coupled type I side 
predicts a bound state
(F1, anti-F1) in the strongly coupled Heterotic theory. 
S-duality determines that each of these
F1's transforms in the spinor representation of SO(32), since DN
open D-strings contribute with these quantum numbers, and that the whole
system is unstable due to the presence of a tachyonic mode in the open
D-strings stretched between the two F1's. S-duality determines as well
that if the system is compactified a state transforming as a spinor
under SO(32) should emerge after the condensation of the tachyonic mode.
This description at strong coupling  in terms of an 
(F1, anti-F1) bound state is also natural from another point
of view. In the Heterotic theory the non-BPS states in the spinorial
representation of SO(32) correspond to unwrapped fundamental strings,
i.e. to strings not charged with respect to the NS-NS 2-form potential.
This charge cancellation is simply achieved by the F1, anti-F1
superposition. 

It is also interesting to point out that 
the Heterotic fundamental string can arise as a bound
state of a pair of NS5, anti-NS5 branes, since this is the S-dual 
configuration of the D1=(D5, anti-D5) bound state in type I \cite{sena}.
This is implied by S-duality, and can be read as
well directly from the NS5-brane effective action truncated
to a Heterotic background, in particular from the term \cite{EJL}:
$\int_{R^{5+1}}{\rm Tr}\left[ B^{(2)}\wedge {\tilde F}\wedge {\tilde F}
\right]$. 
In turn, the NS5-brane is obtained as a bound state of four NS9,
anti-NS9 pairs of branes. This is S-dual to the D5=4(D9, anti-D9)
configuration in type I, and can also be derived from the 
NS9-brane effective
action (\ref{NS9ac}) truncated to a Heterotic 
background.
Indeed, by S-duality,  four NS9, anti-NS9 are characterised by an  
SO(4) $\times$ SO(4) gauge symmetry. One of the SU(2) of the
SO(4) Chan-Paton group, characterising, say, the four anti-NS9 branes, 
is used in the construction of the NS5-brane as an instanton 
configuration and the other SU(2) survives as its gauge symmetry.
The Heterotic 5-brane contains as well SO(32) currents that arise 
from open D-strings with DN
boundary conditions, i.e. with one end on an NS9-brane
\cite{Hull2}. Therefore, each brane of the
NS5, anti-NS5 pair has an SU(2) $\times$ SO(32) gauge
structure, from which the SU(2)
is used in the instanton construction of the Heterotic F-string and the SO(32)
group survives as its gauge structure. 
 
\section{A non-BPS M10-brane}

In this section we briefly discuss how to construct a 
non-BPS M10-brane \cite{HL3}.
The conjectured M10-brane is an unstable
spacetime-filling brane in massive eleven dimensional supergravity.
This brane is constructed
such that the BPS M9-brane is obtained after the tachyonic mode of
an M2-brane ending on it condenses and such that it gives the 
non-BPS D9-brane
of type IIA upon reduction.

The study of 9-branes and 10-branes in M-theory should be carried out in the 
context of massive eleven dimensional supergravity, given that the
BPS M9-brane couples magnetically to the mass. It is well-known 
that a fully eleven dimensional Lorentz invariant massive 
supergravity cannot be constructed \cite{BDHS}. Nevertheless, the massive
Type IIA supergravity of Romans \cite{Romans} can be derived from
eleven dimensions if the condition of eleven dimensional Lorentz
invariance is relaxed and one assumes an isometric
eleventh direction that is gauged in the supergravity action
\cite{BLO}. Accordingly, the branes living
in this massive eleven dimensional background are described by
worldvolume effective actions where the Killing direction, characterised
by a Killing vector ${\hat k}$, is 
gauged. The M9-brane effective action has been constructed
in \cite{Mm}, where it is shown that its worldvolume field 
strength is a 
two-form ${\hat {\cal F}}$ characterising an M2, wrapped on the
Killing direction, ending on the M9. 
The worldvolume theory of the M10 will also contain this 
${\hat {\cal F}}$ field but will contain in addition a real
tachyon field ${\hat T}$ such that upon double dimensional 
reduction the type IIA non-BPS D9-brane \cite{kteo2} is reproduced. 
Under the two hypothesis made, one can thus construct the action
of a non-BPS M10-brane. Here we just present the result. The action is: 
$S^{({\rm M10})}=S^{({\rm M10})}_{{\rm DBI}}+
S^{({\rm M10})}_{{\rm WZ}}$, where the DBI part is given by:

\begin{equation}
\label{M10DBI}
S^{({\rm M10})}_{{\rm DBI}} =-\int_{R^{9+1}} |{\hat k}|^3 
\sqrt{|{\rm det} \left( {\hat \Pi}+ |{\hat k}|^{-1}  
{\hat {\cal F}} \right)|}
\,{\hat R}({\hat T},\partial {\hat T},\dots)\, .
\end{equation}

\noindent Here ${\hat \Pi}$ is the pull-back of the spacetime
metric:

\begin{equation}
{\hat \Pi} = D {\hat X}^{\hat \mu} D {\hat X}^{\hat \nu} 
{\hat g}_{{\hat \mu}{\hat \nu}} = 
\partial {\hat X}^{\hat \mu} \partial {\hat X}^{\hat \nu} 
\left( {\hat g}_{{\hat \mu}{\hat \nu}}
+ |{\hat k}|^{-2}{\hat k}_{\hat \mu}{\hat k}_{\hat \nu} \right) \, ,
\end{equation}
\noindent and ${\hat R}$ is some function of the tachyon field vanishing
at the minimum of the tachyon potential, where the action then 
vanishes identically and  describes a configuration  indistinguishable 
from the vacuum. For vanishing tachyon field ${\hat R}$ 
gives a constant. For a tachyonic kink configuration in the zero size limit
${\hat R}(x)\sim C \delta(x-x_0)$, and (\ref{M10DBI})
reduces to the DBI part of the effective action of a BPS 
M9-brane.
\noindent The Wess-Zumino part is given by the following expression: 

\begin{eqnarray}
\label{M10WZ}
&&S^{(M10)}_{\rm WZ}=\int_{R^{9+1}}\left[ i_{\hat k}{\hat B}^{(10)}+
i_{\hat k}{\hat N}^{(8)}\wedge {\hat {\cal F}}+i_{\hat k}{\hat {\tilde C}}
\wedge {\hat {\cal F}}\wedge {\hat {\cal F}}+ 
{\hat C}_{{\hat \mu}_1{\hat \mu}_2{\hat \mu}_3}.\right. \\ 
&&\left. .D{\hat X}^{{\hat \mu}_1}D{\hat X}^{{\hat \mu}_2}
D{\hat X}^{{\hat \mu}_3}\wedge
{\hat {\cal F}}\wedge {\hat {\cal F}}\wedge {\hat {\cal F}}+
{\hat A}\wedge {\hat {\cal F}}\wedge {\hat {\cal F}}
\wedge {\hat {\cal F}}\wedge {\hat {\cal F}}+\dots \right]\wedge d{\hat T}\, ,
\nonumber
\end{eqnarray}
where ${\hat B}^{(10)}$ is the spacetime field
electric-magnetic dual to the
mass and ${\hat N}^{(8)}$ is the Kaluza-Klein monopole charge.

Having this spacetime-filling M10-brane it is possible to describe
brane descent constructions of all M-branes. We refer the 
interested reader to \cite{HL3}.

\subsection*{Acknowledgements}

L.~H. would like to acknowledge the support of the European Commission 
RTN programme RTN1-1999-00116.

\end{document}